\documentclass[superscriptaddress, aps, pra, twocolumn, nofootinbib]{revtex4-2}

\usepackage{dsfont}
\usepackage{upgreek}
\usepackage{amsfonts}
\usepackage{amsmath}
\usepackage{amsthm}
\usepackage{amssymb}
\usepackage{graphicx}
\usepackage{dcolumn}
\usepackage{bm}
\usepackage[colorlinks, allcolors=blue]{hyperref}
\usepackage{braket}
\usepackage{url}
\usepackage{enumitem}
\usepackage[dvipsnames]{xcolor}
\usepackage{soul}
\usepackage{parskip}
\usepackage{apptools}
\usepackage{stackrel}
\usepackage{dsfont}
\usepackage{indentfirst}
\usepackage[normalem]{ulem}

\newcommand{\stkout}[1]{\ifmmode\text{\sout{\ensuremath{#1}}}\else\sout{#1}\fi}

\usepackage{tikz, pgfopts}
\usepackage{tikz-3dplot}
\usetikzlibrary{positioning}

\definecolor{Red}{rgb}{0.9,0,0}
\definecolor{Blue}{rgb}{0,0,0.9}
 
\newcommand{\kp}{\ket{\psi}}

\newcommand{\C}{\mathds{C}}
\newcommand{\I}{\mathds{I}}

\newtheorem{theorem}{Theorem}[section]

\begin{document}

\title{From Rotations to Unitaries: Reversible Quantum Processes and the Emergence of the $SU(2)-SO(3)$ Homomorphism}

\author{V. G. Valle}
\affiliation{Depto. de F\'isica, ICE, Universidade Federal de Juiz de Fora, MG, Brazil}

\author{B. F. Rizzuti}
\email{brunorizzuti@ufjf.br}
\affiliation{Depto. de F\'isica, ICE, Universidade Federal de Juiz de Fora, MG, Brazil}

\begin{abstract}
\begin{center}
\textbf{Abstract}
\end{center}

We present an operational reconstruction of the well-known two-to-one homomorphism between the groups $SU(2)$ and $SO(3)$, grounded in the physical description of quantum state preparation and evolution. Starting from the connection between vectors in three-dimensional physical space and quantum states of two-level systems, we investigate how reversible transformations—modeled as completely positive and trace-preserving maps—give rise to a correspondence between spatial rotations and unitary operations. Our approach reveals how this group-theoretic structure naturally emerges from physical constraints, particularly the preservation of purity and reversibility in quantum processes. Beyond its theoretical relevance, the construction offers a pedagogically accessible framework for introducing core ideas in quantum mechanics and symmetry groups, making the abstract correspondence between $SU(2)$ and $SO(3)$ tangible through experimentally meaningful procedures.   

\end{abstract}

\maketitle

\section{Introduction}
\label{Sec.1}

Physical processes in quantum theory are described by completely positive and trace-preserving (CPTP) maps. By processes, we intend this term to be as broad as possible. Examples include measurement processes via quantum instruments \cite{leifer_towards_2013}, (ir)reversible time evolution \cite{wolf}, coarse-grained descriptions of blurred or saturated detectors \cite{duarte_emerging_2017}, local operations and classical communication protocols composed of CPTP maps for applications such as entanglement theory \cite{horodecki_quantum_2009}, and the modeling of open quantum systems' dynamics, where CPTP maps are employed to reproduce non-unitary evolution due to decoherence and possible dissipation \cite{marquardt_introduction_2008}. This list can be extended indefinitely, reflecting the ubiquitous role of such maps in quantum theory and beyond --- generalized probability theories, for example \cite{plavala_general_2023}.

To get a sense of why CPTP maps are of interest, it is helpful to break down the name into two parts. Although one may start from convexity, these maps can be extended to transform quantum states into quantum states in a linear fashion in the whole space. Since quantum states are trace-class operators (i.e., positive operators with unit trace), it is natural to require the trace-preserving (TP) condition, ensuring that the unit trace remains unchanged under the action of the map.

Before the CP part, we start with the notion of only positivity, which is central to the mathematical and physical structure of quantum theory. In particular, positive operators are essential in the description of quantum state operators because positivity guarantees that all measurement probabilities are non-negative. Moreover, a density operator must be positive semidefinite to ensure that expectation values of all observables correspond to physically meaningful, real, and non-negative probabilities.

Going a little further, complete positivity is also a fundamental requirement for any physically admissible map. When a system interacts with an external environment, for instance, the joint evolution is described by a unitary transformation acting on the combined Hilbert space. After tracing out the environment, the reduced dynamics of the system are represented by a map acting solely on its state space. To ensure that this reduced map yields a valid quantum state not only for isolated systems but also when the system is part of a larger entangled one, the map must be completely positive. Without complete positivity, the extended action of the channel on entangled states could produce non-physical, non-positive density operators, violating the probabilistic consistency of quantum theory.  A standard example is the transpose operation, which is positive but not completely positive \cite{nielsen2010quantum}. For this reason, we also require the completely positive (CP) part. 

Together, these two conditions---which define what is known as a quantum channel (or physical map)---appear to impose strong restrictions on the invertibility of CPTP maps with inverses that are still physically admissible.  The depolarizing channel provides an example of a CPTP map whose inverse is not positive, highlighting the rarity --- or even contingency --- of finding a CPTP map with a CPTP inverse \cite{nayak2006invertible}.

This question has already been studied in detail \cite{nayak2006invertible}, where the characterization of invertible quantum channels is addressed via their Kraus representation. In essence, a physical map admits a CPTP inverse if and only if the Kraus decomposition of the former contains exactly one Kraus operator. Being a CPTP map, one demands that the Kraus representation is trace-preserving; this translates into the operators, say, $A_a$, with $a$ running in a finite set, adding up to identity, $\sum_a A^*_aA_a = \mathds{I}$. With just one operator in the representation, one concludes that the only operator is, actually, unitary.

In our contribution, we take this analysis a step further by exploring the connection between invertible CPTP maps with physical admissible inverse and group theory. Remarkably, the well-known two-to-one homomorphism between  $SU(2)$ and $SO(3)$ can be recovered in a surprisingly simple way through a preparation procedure involving two-level quantum systems embedded in three-dimensional physical space.

This work is a natural continuation of \cite{valle2024towards}, where the Hopf fibration is given an operational interpretation in terms of state preparation. The result is an intrinsic connection between the physical space, denoted by $\mathds{D} \cong \mathds{R}^3$ and qubits in $\mathds{C}^2$. The authors take Asher Peres's saying 

\textit{quantum phenomena do not occur in a Hilbert space,
they occur in the laboratory} \cite{peres_quantum_2010} 

to emphasize that any two-level quantum system can, in practice, be prepared by selecting appropriate directions in three-dimensional physical space, as encoded operationally by the Bloch representation.

In this work we intentionally separate two distinct uses of the symbol $\mathds{R}^3$: (i) the physical space where laboratory procedures take place, and (ii) the Euclidean vector space that is used as a convenient representation space (for example the ambient space that contains the Bloch sphere). To avoid conflation we adopt an operational construction of physical space following the approach of \cite{gaio_grandezas_2019, rizzuti_operational_2020}: physical space is obtained from experimentally accessible procedures (marking points on rigid reference bodies, defining distances by compasses, forming classes under ``firm unions” of rigid bodies, etc.), which yields a system of reference (SR) and an associated affine space $\mathcal{E}_{SR}$ together with its vector space of displacements $\mathds{D}$. This operational viewpoint emphasizes that points, displacements (vectors), directions and affine structure are defined by procedures and not assumed a priori.

With that operational scaffolding in place we can state the relation between Bloch vectors and the laboratory: Bloch vectors are elements of an abstract three-dimensional real vector space (the space in which the Bloch sphere $S^2$ is embedded), and, when a specific physical implementation is chosen, these vectors acquire an operational meaning as directions that label preparation procedures (for instance the orientation of a Stern–Gerlach magnetic field \cite{grossi2023one} or the settings of an $SU(2)$ polarization gadget \cite{valle2024towards}). The Hopf fibration provides the mathematical bridge: indistinguishable pure state rays in the normalized Hilbert sphere $S^3 \subset \mathds{C}^2$ are projected onto points of $S^2\subset \mathds{R}^3$, so that each point on the Bloch sphere corresponds operationally to the family of laboratory preparations that produce that same state up to a global phase. The operational recipe for realizing the spherical coordinates 
$(\theta, \varphi)$ is therefore apparatus-dependent (\textit{e.g.} SG magnet orientation or the Euler-angle settings of the universal $SU(2)$ polarization gadget), but the Hopf map guarantees a system-independent identification between the equivalence class of preparations and a Bloch direction.

\begin{figure}[ht]
    \centering
    \caption{Diagram representing the allegedly relation between the groups $SO(3)$ and $SU(2)$.}
    \includegraphics[scale=0.34]{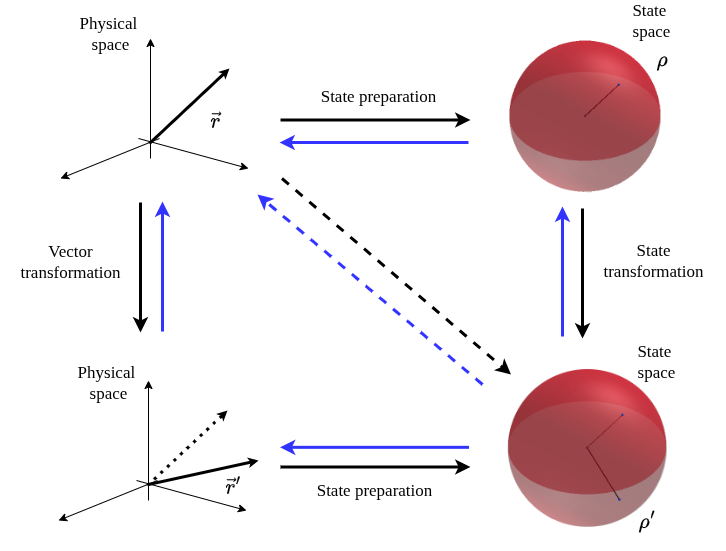}
    \label{fig:1}
\end{figure}

Building on this, we go further to investigate transformations between vectors in $\mathds{D}$ (which give rise to quantum states in the sense explained above) and the corresponding allowed state transformations in the space of states. The sequence of these transformations is expected to make the diagram in Fig.~\ref{fig:1} commute.

On the one hand, an allowed transformation of a vector in physical space is represented by a rotation, which in turn induces a unitary transformation of the corresponding quantum state. This gives rise to the direction 
$SO(3)\rightarrow SU(2)$ in the homomorphism we aim to recover. It corresponds to the path represented by the black horizontal arrows in the diagram shown in Fig.~\ref{fig:1}. On the other hand, the reverse direction, $SU(2) \rightarrow SO(3)$, is obtained by requiring a reversible physical transformation of a quantum state. As previously discussed, the only permissible process of this kind is a unitary operation, which translates into a rotation in physical space. This step is illustrated by the blue arrows in the diagram.

To construct both directions in the diagram, we proceed as follows. In the next section, we show how a rotation in physical space unambiguously induces a unitary transformation on the quantum state associated with the corresponding vector. Section~\ref{Sec.3} provides a brief review of CPTP maps, with a focus on their Kraus representation. In particular, we discuss a characterization theorem for invertible CPTP maps with physically admissible inverses. Building on this result, in Section~\ref{Sec.4}, we start from a unitary representing a reversible quantum process and reconstruct the associated rotation, thereby recovering the group two-to-one homomorphism we set out to rediscover. Finally, Section~\ref{Sec.5} is left for conclusions and some perspectives.

\section{From physical spaces to state spaces: $SO(3) \rightarrow SU(2)$} \label{Sec.2}

Our starting point is the connection between physical space and the space of quantum states. From there, we begin to explore the relationship between transformations on each side ---specifically, the correspondence between rotations and unitary transformations.

Given a unit vector (called the Bloch vector) belonging to what is called the Bloch sphere $S^2$
\begin{equation}
    \vec{r} = (\sin\theta \cos \varphi, \sin \theta \sin\varphi, \cos \theta),
\end{equation}
with $\theta \in [0, \pi]$ and $\varphi \in [0, 2\pi)$ in the physical space $\mathds{D} \cong \mathds{R}^3$, one constructs the pure state in $\mathds{C}^2$, or strictly speaking, in the projective space $\mathds{CP}^1$,
\begin{equation}\label{5.1}
    \kp = \cos\frac{\theta}{2} \ket{0}+ e^{i \varphi}\sin\frac{\theta}{2} \ket{1}
\end{equation}

The parameters $\theta$ and $\varphi$ may have a concrete physical interpretation. For example, in \cite{grossi2023one} it is shown that they indicate the direction of a magnetic field in the Stern-Gerlach experiment, preparing  a bean of spin-up particles in that direction. Capitalizing in the Fig. \ref{fig:1}, analogous to the isometry of $SU(2)$ and $SO(3)$, there is the isometry of the complex projective space $\mathds{CP}^1$ --- the space of pure states of a qubit, modulo the global phase --- with the Bloch sphere $S^2$ \cite{bengtsson2017geometry}. We restrict ourselves to the former.
 
The corresponding density operator related to \eqref{5.1} is given by
\begin{equation}\label{5.2}
    \rho = \ket{\psi}\bra{\psi}  = \frac{1}{2}(\mathds{I} + \vec{r} \cdot \vec{\sigma})
\end{equation}
where $\vec{\sigma} = (\sigma_x, \sigma_y, \sigma_z)$ are the Pauli matrices. 

The right-hand-side of \eqref{5.2} is not restricted to this particular example of a pure state. Actually, it represents the most general density operator in $L(\mathds{C}^2)$, the space of linear operators over $\mathds{C}^2$. 

Let $\xi$ be an arbitrary density operator. Knowing that 
\begin{enumerate}
    \item $Tr \,\xi^2 \leq 1$

\item $\xi$ is pure if and only if $Tr \,\xi^2 = 1,$ 
\end{enumerate}
we find that $\rho$ in Eq. \eqref{5.2} is pure if and only if $\Vert \vec{r} \Vert=1$ and mixed when $\Vert \vec{r} \Vert < 1$. 

With this result in hand, we turn to transformations in physical space that preserve the nature of the quantum state---whether pure or mixed. Such transformations must be isometries, and can be described as rotations around a given axis $\hat{n}$ by an angle $\alpha$, denoted $\mathcal{R}(\hat{n},\alpha)$. 

With all that has being said, the next theorem shows how a transformation in a vector in the physical space induces a change in the corresponding state on/in the Bloch sphere; see, for example \cite{gamel_entangled_2016}. 

\begin{theorem}\label{theorem.0}
    Given a general rotation element $\mathcal{R}(\hat{n}, \alpha)\in SO(3)$. The state $\rho_{\vec{r}'}$, described by the vector  $\vec{r}\,' = \mathcal{R}(\hat{n}, \alpha) \vec{r}$, can be calculated by applying a matrix $\mathcal{U}(\hat{n}, \alpha)$ $\in SU(2)$ on the state $\rho_{\vec{r}}$, described by the original vector $\vec{r}$, in such a way that $\rho_{\vec{r}'} = \mathcal{U}(\hat{n}, \alpha) \rho_{\vec{r}} \,\mathcal{U}^*(\hat{n}, \alpha)$.
\end{theorem}
\textit{\underline{Proof:}}

A general rotation element of $SO(3)$ is given by the Rodrigues's formula  \cite{mukunda2010hamilton}
\begin{align}
 \mathcal{R}_{jk}(\hat{n}, \alpha) &= \delta_{jk}\cos \alpha + n_j n_k (1-\cos \alpha) - \sum^3_{l=1}\varepsilon_{jkl}n_l
 \sin \alpha \nonumber \\ \label{4}
 &= \left ( e^{-i \alpha \hat{n}\cdot \vec{\tau}} \right )_{jk}
\end{align}
and the group generators $\vec{\tau}$ are given by $(\tau_l)_{jk} = -i \varepsilon_{jkl}$ \cite{tung_group_1985}. From now on, Latin letters ($i, j, k, ...$), as customary, assume the values $1, 2, 3$. 

Now, if $\vec{r}\,' = \mathcal{R}(\hat{n},\alpha) \vec{r}$, then the rotated vector $\vec{r}\,'$ gives rise to a state $\rho_{\vec{r}'}$ that we seek to determine. We have
\begin{align}
    \rho_{\vec{r}'} &= \frac{1}{2} \left ( \mathds{I}+ \vec{r}\,' \cdot \vec{\sigma} \right ) = \frac{1}{2} \left ( \mathds{I}+ (\mathcal{R}\vec{r}) \cdot \vec{\sigma} \right ) \nonumber \\ 
    &= \frac{1}{2} [\mathds{I} +\cos \alpha (\vec{r}\cdot \vec{\sigma})+ (1- \cos \alpha )(\hat{n} \cdot \vec{r})(\hat{n} \cdot \vec{\sigma}) \nonumber \\ &-\sin \alpha (\vec{r} \times \hat{n}) \cdot \vec{\sigma}] \label{5}.
\end{align} 
In the expression above, we have used the explicit action of \eqref{4} on the vector $\vec{r}$, that is,
\begin{equation}
 \vec{r}\,' = \cos \alpha \vec{r} + (1-\cos\alpha)(\hat{n}\cdot \vec{r})\hat{n} - \sin\alpha \,\vec{r} \times \hat{n}.   
 \nonumber
\end{equation}

Using the trigonometrical identities
\[
\cos\alpha = \cos^2{\frac{\alpha}{2}}-\sin^2{\frac{\alpha}{2}}, \quad \,\, \sin\alpha=2\sin{\frac{\alpha}{2}}\cos\frac{\alpha}{2},
\]
and 
\[
1-\cos\alpha = 2\sin^2{\frac{\alpha}{2}}
\]
one finds that
\begin{align}
    &\vec{r}\,'\cdot\vec\sigma = \left(\cos^2{\frac{\alpha}{2}}-\sin^2{\frac{\alpha}{2}}\right)(\vec{r}\cdot\vec{\sigma}) -\nonumber \\ &- 2\sin{\frac{\alpha}{2}}\cos{\frac{\alpha}{2}}(\vec{r}\times\hat{n})\cdot\vec\sigma + 2\sin^2{\frac{\alpha}{2}}(\hat{n}\cdot\vec{r})(\hat{n}\cdot\vec{\sigma}). \nonumber
\end{align}
Rearranging terms, we have
\begin{align}
    \vec{r}\,'\cdot\vec\sigma&= \cos^2{\frac{\alpha}{2}}(\vec{r}\cdot\vec{\sigma})- 2\sin{\frac{\alpha}{2}}\cos{\frac{\alpha}{2}}(\vec{r}\times\hat{n})\cdot\vec\sigma \nonumber\\ &+ \sin^2{\frac{\alpha}{2}}(2(\hat{n}\cdot\vec{r})(\vec{r}\cdot\vec{\sigma})-(\vec{r}\cdot\vec{\sigma})).\nonumber
\end{align}

From $(\vec{r}\cdot\vec\sigma)(\vec{n}\cdot\vec\sigma) = (\vec{r}\cdot\hat{n})\mathds{I}+i(\vec{r}\times\hat{n})\cdot\vec\sigma$, the Pauli product identity, we can extract two identities, namely,
\begin{align}
(\hat{n}\cdot\vec{\sigma})(\vec{r}\cdot\vec{\sigma})(\hat{n}\cdot\vec{\sigma}) = 2(\hat{n}\cdot\vec{r})(\vec{r}\cdot\vec{\sigma})-(\vec{r}\cdot\vec{\sigma}); \nonumber \\(\vec{r}\cdot\vec{\sigma})(\hat{n}\cdot\vec{\sigma})-(\hat{n}\cdot\vec{\sigma})(\vec{r}\cdot\vec{\sigma}) = 2i(\vec{r}\times\hat{n})\cdot\vec{\sigma}. \nonumber
\end{align}
Knowing this, our $\vec{r}\,'\cdot\vec\sigma$ term than becomes
\begin{align}
     \vec{r}\,'\cdot\vec\sigma&= \cos^2{\frac{\alpha}{2}}(\vec{r}\cdot\vec{\sigma}) - i\cos{\frac{\alpha}{2}}\sin{\frac{\alpha}{2}}(\hat{n}\cdot\vec{\sigma})(\vec{r}\cdot\vec{\sigma})+ \nonumber \\
     &+i\cos{\frac{\alpha}{2}}\sin{\frac{\alpha}{2}}(\vec{r}\cdot\vec{\sigma})(\hat{n}\cdot\vec{\sigma})+ \nonumber \\
     &+\sin^2{\frac{\alpha}{2}}(\hat{n}\cdot\vec{\sigma})(\vec{r}\cdot\vec{\sigma})(\hat{n}\cdot\vec{\sigma}),\nonumber
\end{align}
and finally, with $\mathcal{U} = \cos{\frac{\alpha}{2}}\mathds{I}-i(\hat{n}\cdot\vec{\sigma})\sin{\frac{\alpha}{2}}$, we can rearrange the expression above to
\begin{equation}\label{11}
    \vec{r}\,'\cdot\vec\sigma= \mathcal{U} \ (\vec{r}\cdot\vec{\sigma})\ \mathcal{U}^*.
\end{equation}

The expression \eqref{11} allows us to rewrite our $\rho_{\vec{r}'}$ from equation \eqref{5} as
\begin{equation}
\label{UphiU}
   \rho_{\vec{r}'} = \mathcal{U}(\hat{n}, \alpha) \rho_{\vec{r}} \, \mathcal{U}^*(\hat{n}, \alpha)
\end{equation} 
where $\mathcal{U}(\hat{n}, \alpha)$ is the unitary matrix
\begin{equation}\label{7}
    \mathcal{U}(\hat{n}, \alpha) = \cos \frac{\alpha}{2} \mathds{I} - i \sin \frac{\alpha}{2} (\hat{n} \cdot \vec{\sigma})= e^{-i \frac{\alpha}{2}\hat{n}\cdot \vec{\sigma}},
\end{equation}
with $\det \mathcal{U}(\hat{n}, \alpha)=1$, thus completing the proof. $\qedsymbol{}$ 

The result above shows that each rotation $\vec{r}\,' = \mathcal{R}(\hat{n},\alpha) \vec{r}$, with $\mathcal{R}(\hat{n},\alpha) = e^{-i \alpha \hat{n}\cdot \vec{\tau}}$ induces  a state transformation $\rho_{\vec{r}'}= \mathcal{U}(\hat{n}, \alpha) \rho_{\vec{r}} \, \mathcal{U}^*(\hat{n}, \alpha)$ with 
$\mathcal{U}(\hat{n}, \alpha) = e^{-i \frac{\alpha}{2}\hat{n}\cdot \vec{\sigma}}$. Hence we have just established the mapping $\phi: SO(3) \rightarrow SU(2)$, with $\phi(\mathcal{R}(\hat{n},\alpha)) = \mathcal{U}(\hat{n}, \alpha)$. It is important to emphasize a peculiarity of this mapping. Since in \eqref{UphiU} the unitary operator $\mathcal{U}$ and its conjugate appear jointly, the elements $\mathcal{U}$ and $-\mathcal{U}$ yield the same physical result. Therefore, \eqref{7} does not, strictly speaking, define a mapping from spatial rotations to elements of $SU(2)$, but rather from the set of parameters $(\vec{n}, \alpha)$ to $SU(2)$. Additional remarks concerning this behavior will be provided in Section~\ref{Sec.4}, specifying the double-cover structure of $SU(2)$ over $SO(3)$.

Our next step will be focused on constructing the inverse of $\phi$. To that end, we will make a short digression on CPTP maps. 

\section{Brief review on CPTP maps}\label{Sec.3}

Let us consider the spaces $L(\mathds{C}^n)$ and $L(\mathds{C}^m)$ of linear transformations on $\mathds{C}^n$ and $\mathds{C}^m$, respectively. A map
\begin{equation}
    \Phi:    L(\mathds{C}^n) \to L(\mathds{C}^m)
\end{equation}
is CPTP whenever $Tr(X) = Tr(\Phi(X))$, for all $X \in L(\mathds{C}^n)$ and 
$(\Phi \otimes \mathds{I}_{L(\mathds{C}^q)})(P) \in Pos (\mathds{C}^m \otimes \mathds{C}^p)$ for all $P \in Pos (\mathds{C}^n \otimes \mathds{C}^q)$ and for every choice of $\mathds{C}^q$. As usual, $Pos(\mathds{C}^\aleph)$ is the set of positive semidefinite operators on  $\mathds{C}^\aleph$, for some positive integer $\aleph$.

There are several ways to represent $\Phi$, which include the natural one, Stinespring, Choi-Jamiołkowski and Kraus representations, all of them being equivalent \cite{Watrous_2018}. Given our goal of relating reversible CPTP maps to group theory, we will use the Kraus representation, which is given by
\begin{equation}\label{8.1}
    \Phi(\rho) = \sum_{a\in \Gamma}A_a \rho A^*_a, \,\, \rho \in L(\mathds{C}^n)
\end{equation}
for a finite non-empty set $\Gamma$ and $\{ A_a \, \vert \, a \in \Gamma \} \subset L(\mathds{C}^n, \mathds{C}^m)$. The maps $A_a$ are called Kraus operators. 

The map \eqref{8.1} is CP and the TP is guaranteed asking for \cite{Watrous_2018}
\begin{equation}\label{8.11}
    \sum_{a\in\Gamma}A^*_aA_a = \mathds{I}_{n}.
\end{equation}

The formulation of conditions that make $\Phi$ an invertible map has already been studied. For example, the Kadison's theorem states that if a map $\Phi$ from density operators $\mathcal{D}(\mathds{C}^n)$ to $\mathcal{D}(\mathds{C}^n)$ is one-to-one, onto and convexity-preserving, then 
\begin{equation}\label{9.1}
    \Phi(\rho) = \mathcal{U}\rho\, \mathcal{U}^*,
\end{equation}
where $\mathcal{U}$ is either unitary or anti-unitary \cite{bengtsson2017geometry}. The structure of \eqref{9.1} is a hint that a bijective quantum channel admits only one Kraus operator. Indeed, Theorem 2.1 in \cite{nayak2006invertible} states that whenever a CPTP map admits a CPTP inverse, its Kraus representation must be of the form given in \eqref{9.1}. With precise details,
\\
\begin{theorem}\label{theorem.1}
    Let $\Phi: L(\mathds{C}^n) \rightarrow L(\mathds{C}^m)$ be a CPTP map. Suppose that $\Phi$ admits an inverse $\Phi^{-1}$ which is also CPTP. Then there exists a density matrix $\omega \in L(\C^{\lfloor m/n\rfloor})$ and a unitary $\mathcal{U}$ on $\mathds{C}^m$ such that 
    $\Phi(\rho) = \mathcal{U}(\rho\otimes\omega) \, \mathcal{U}^*$ and $\Phi^{-1}$ corresponds to applying $\mathcal{U}^*$ and tracing out the $\lfloor m/n\rfloor$ dimensional ancilla. Conversely, if $\Phi$ is a CPTP map whose Kraus representation has only one operator, say $\Phi(\cdot) = \mathcal{U}(\cdot) \, \mathcal{U}^*$, then $\Phi$ is invertible and $\Phi^{-1}$ is also CPTP. 
\end{theorem}

Although somewhat redundant, we state and prove a corollary of the above theorem for completeness in our manuscript, in the case where $\Phi: L(\mathds{C}^n) \rightarrow L(\mathds{C}^n)$. We follow the same lines of reasoning as in \cite{nayak2006invertible}. Because the proof is somehow constructive, we also present it here for didactic purposes --- see the Appendix on Sec. \ref{sec.appendix}. 

This theorem serves as the foundation for constructing invertible physical processes that exhibit a group structure. It enables us to obtain a rotation in physical space starting from a unitary transformation, as we shall see in the next section.

\section{From state spaces to physical spaces: $SU(2) \rightarrow SO(3)$}\label{Sec.4}

Now, suppose we have an invertible physical process evolving a state $\rho$ on (or in) the Bloch sphere. Due to the theorem \ref{theorem.1}, its Kraus representation is given by
\begin{equation}\label{evolution}
    \Phi(\rho)= \mathcal{U}\rho \, \mathcal{U}^*,
\end{equation}
with $\mathcal{U}^*\mathcal{U} = \mathds{I}_{\mathds{C}^2}$. This restriction of unitarity implies $\det \mathcal{U} = e^{i\lambda}$, $\lambda \in \mathds{R}$. Because we are aiming the group $SU(2)$, we can consider only unitary matrices with determinant equals one. In fact, if $\mathcal{U}$ is such that $\det \mathcal{U} = e^{i\lambda}$, define
\[
\mathcal{W}:= \frac{\mathcal{U}}{\sqrt{\det\mathcal{U}}}.
\]
Therefore, not only $\det \mathcal{W}=1$ but also $\mathcal{W}$ defines the same channel as $\mathcal{U}$,
\begin{equation}
\mathcal{W}\rho\mathcal{W^*} = \frac{\mathcal{U}}{\sqrt{\det\mathcal{U}}}\rho\frac{\mathcal{U^*}}{\sqrt{\overline{\det\mathcal{U}}}} = \frac{1}{\vert \det\mathcal{U}\vert^2}\mathcal{U}\rho\mathcal{U}^*=\mathcal{U}\rho\mathcal{U}^* \nonumber
\end{equation}

Hence, with no loss of generality, it is enough to consider only $\mathcal{U} \in SU(2)$ and we parametrize the group elements according to the section \ref{Sec.2}: $\mathcal{U}(\hat{n}, \alpha) = e^{-i \frac{\alpha}{2}\hat{n}\cdot \vec{\sigma}}$. The transition from \eqref{5} to \eqref{UphiU} can be reverted. That is, starting from the right hand side of \eqref{11}, all the trigonometrical identities can be used in the opposite order of the ones taken in the steps from $SO(3)$ to $SU(2)$. Thus, $\mathcal{U}(\hat{n}, \alpha)$ induces the rotation $\mathcal{R}(\hat{n}, \alpha)$ given by \eqref{4}. We are finally led to $\phi^{-1}(\mathcal{U}(\hat{n}, \alpha)) = \mathcal{R}(\hat{n}, \alpha)$. 

 $\phi^{-1}$ is not a true inverse, because $SU(2)$ is a double-cover to $SO(3)$, a known result in group theory \cite{hall_lie_2015}. In other words, two elements of $SU(2)$ are mapped to one element in $SO(3)$. It becomes clear if we obtain the matrix element of $SO(3)$ in terms of the unitary element in $SU(2)$, which will be our next step. 

We begin by pointing out the bijection between the Lie algebra $su(2)$ of the group $SU(2)$ and the physical space $\mathds{D}$. Explicitly, $su(2)$ is the set of complex traceless matrices $u$ such that $u^* = -u$, equipped with the commutator $[u_1,u_2] = u_1 u_2 - u_2u_1$, for arbitrary elements $u_1$ and $u_2$ in the algebra.

$su(2)$ is a three-dimensional space spanned by the Pauli matrices. An arbitrary element $u\in su(2)$ is given by
\begin{equation}
    u = i \vec{r}\cdot \vec{\sigma},
\end{equation}
which makes explicit the one-to-one so stated relation  $su(2) \cong \mathds{D}$,
\begin{align}
    \Psi: su(2) &\rightarrow \mathds{D} \cr
    i \vec{r}\cdot \vec{\sigma} &\mapsto \vec{r}.
\end{align}

One notes that $\det u = \Vert \vec{r} \Vert$. Because we are looking for a transformation (rotation) that is an isometry, our suggestion is to set $u \mapsto u' = \mathcal{U} u \, \mathcal{U}^*$, with $\mathcal{U} \in SU(2)$. In this case, 
$\det u' = \det u \Leftrightarrow \Vert \vec{r}\,' \Vert = \Vert \vec{r} \Vert$, inducing, then, a transformation in the vector $\Psi(i \vec{r}\cdot \vec{\sigma}) = \vec{r}$ that preserves its norm. The transformation in $su(2)$ is just a group homomorphism
\begin{align}
    T: SU(2) &\rightarrow L(su(2)) \cr
    \mathcal{U} &\mapsto T_{\mathcal{U}}: su(2) \rightarrow su(2).  
\end{align}
where $u   \stackbin{T_{\mathcal{U}}}{\longmapsto} \mathcal{U}u \, \mathcal{U}^*$. 
We observe that $T_{\mathcal{U}}$ is linear and for $\mathcal{U}, \mathcal{V} \in SU(2)$ 
\[
T_{\mathcal{U}}T_{\mathcal{V}} = T_{\mathcal{U}\mathcal{V}}
\]
as expected for a linear group representation \cite{rocha_transformacoes_2013}. 

Well, if $\mathcal{U}$ induces a transformation in $su(2)$ that seen by the eyes of $\mathds{D}$ does not change the norm of a vector, there is no other choice but to connect $\mathcal{U}$ with a rotation in $SO(3)$. We name it $\mathcal{R}$. Thus, we have
\begin{equation}\label{13.1}
    u' = T_\mathcal{U}u = \mathcal{U}(i \vec{r} \cdot \vec{\sigma})\,\mathcal{U}^*= i \vec{r}\,' \cdot \vec{\sigma} = i (\mathcal{R} \vec{r})\cdot \vec{\sigma},
\end{equation}
which is equivalent to
\begin{equation}\label{13.2}
    \sum_j \mathcal{U} x_j \sigma_j\,\mathcal{U}^* = \sum_{l,j} \mathcal{R}_{lj}x_j \sigma_l.
\end{equation}
The standard notation $\vec{r} = (x_1, x_2, x_3)$ has been used. Knowing that $Tr(\sigma_k \sigma_j) = 2 \delta_{kj}$, we multiply both sides of \eqref{13.2} by $\sigma_k$ and trace them out, finding
\begin{equation}
   \sum_j Tr(\mathcal{U}\sigma_j \,\mathcal{U}^*\sigma_k)x_j = 2 \sum_j \mathcal{R}_{kj}x_j, \,\, \forall \vec{r}. 
\end{equation}
The arbitrariness in $\vec{r}$ implies
\begin{equation}\label{30}
    \mathcal{R}_{kj} = \frac{1}{2}Tr(\mathcal{U}\sigma_j \,\mathcal{U}^*\sigma_k).
\end{equation}
All the above computations are schematically expressed in Fig. \ref{fig2}.
\begin{figure}[h]
\begin{center}
    \begin{tikzpicture}[scale=1.0]
        \shade[ball color=white!10!black,opacity=0.5] (0,0) ellipse (40pt and 20pt);
        \shade[ball color=white!10!black,opacity=0.5] (5,0) ellipse (40pt and 20pt);
        \shade[ball color=white!10!black,opacity=0.5] (0,4) ellipse (40pt and 20pt);
        \shade[ball color=white!10!black,opacity=0.51] (5,4) ellipse (40pt and 20pt);
        \shade[ball color=white!10!black,opacity=0.3] (0,2) ellipse (34pt and 16pt);
        \shade[ball color=white!10!black,opacity=0.3] (5,2) ellipse (34pt and 16pt);
        \node (a) at (0,4){};
        \node (b) at (5,4){};
        \node (u) at (-0.5,0){};
        \node (u') at (0.5,0){};
        \node (e) at (4.5,0){};
        \node (e') at (5.5,0){};
        \draw[->, ultra thick, black, bend left] (a) to (b);
        \draw[->, very thick, black] (u) to (u');
        \draw[->, very thick, black] (e) to (e');
        \draw[->, very thick, black, bend right] (u) to (e);
        \draw[->, very thick, black, bend right] (u') to (e');
        \draw node at (0,5) {$SU(2)$};
        \draw node at (5,5) {$SO(3)$};
        \draw node at (1,0.8) {$su(2)$};
        \draw node at (5.8,0.8) {$\mathds{D}$};
        \draw[ultra thick][->][black] (0,4) --   (0,2.2);
        \draw[ultra thick][->][black] (0,1.8) --   (0,0);
        \draw[ultra thick][->][black] (5,4) --   (5,2.2);
        \draw[ultra thick][->][black] (5,1.8) --   (5,0);
        \draw node at (0,2) {$T_{\mathcal{U}}$};
        \draw node at (5,2) {$\mathcal{R}$};
        \draw node at (1,2.7) {$L(su(2))$};
        \draw node at (5.8,2.7) {$L(\mathds{D})$};
        \filldraw[black] (0,4) circle (0.05cm) node [anchor=south]{$\mathcal{U}$};
        \filldraw[black] (5,4) circle (0.05cm) node [anchor=south]{$\mathcal{R}$};
        \filldraw[black] (-0.5,0) circle (0.05cm) node [anchor=south]{$u$};
        \filldraw[black] (0.5,0) circle (0.05cm) node [anchor=south]{$u'$};
        \filldraw[black] (5.5,0) circle (0.05cm) node [anchor=south]{$\vec{r}\,'$};
        \filldraw[black] (4.5,0) circle (0.05cm) node [anchor=south]{$\vec{r}$};
    \end{tikzpicture}
    \end{center}
\caption{Picture of $SU(2)$ and $SO(3)$ via the  $su(2) \cong \mathds{D}$ bijection.}
\label{fig2}
\end{figure}

Eq. \eqref{30} shows that both $\mathcal{U}$ and $-\mathcal{U}$ are mapped to the same element in $SO(3)$, resulting in the two-to-one homomorphism, as previously stated. 

The computations above explores the bijection between two (apparently) distinct vector spaces to obtain the advocated homomorphism. One could, however, reach the same result with only a group theoretical standpoint. To see this, firstly we point out that Eq. \eqref{7} indicates that
\begin{align}
-\mathcal{U}(\hat{n},\alpha) &= -\cos\frac{\alpha}{2}\mathds{I} - i\sin \frac{\alpha}{2}(-\hat{n}\cdot\vec{\sigma}) \nonumber \\
&=\cos \left (\frac{2\pi - \alpha}{2}\right )\mathds{I}-i\sin\left ( \frac{2\pi-\alpha}{2}\right )(-\hat{n}\cdot\vec{\sigma}) \nonumber \\
&= \mathcal{U}(-\hat{n},2\pi-\alpha).\nonumber
\end{align}

On the other hand,
\begin{equation}
\mathcal{R}(-\hat{n},2\pi-\alpha) = \mathcal{R}(\hat{n},\alpha),\nonumber
\end{equation}
which can be concluded by geometric grounds: these formulae shows that rotating around the opposite axis $-\hat{n}$ produces the same effect as rotating around the original axis $\hat{n}$, but at the angle $2\pi-\alpha$. This simple observation also makes it clear that two distinct unitaries, namely $\mathcal{U}$ and $-\mathcal{U}$ in $SU(2)$, are mapped to the same rotation in $SO(3)$.

It is instructive to show that $\mathcal{R}_{kj}$ is indeed an orthogonal matrix. For that, we express \eqref{13.1} in terms of the basis elements $\sigma_i$'s,
\begin{equation}
    \mathcal{U}\sigma_j \, \mathcal{U^*} = \sum_k\mathcal{R}_{kj}\sigma_k.
\end{equation}
Combining two such expressions gives
\begin{equation}
    \mathcal{U}\sigma_j \, \mathcal{U^*}\mathcal{U}\sigma_l \, \mathcal{U^*} = \sum_{k,m}\mathcal{R}_{kj}\sigma_k\mathcal{R}_{ml}\sigma_m.
\end{equation}
Knowing that (i) $\mathcal{U^*} \mathcal{U} = \mathds{I}$; (ii) $Tr(AB) = Tr(BA)$, and, once again, (iii) $Tr(\sigma_i \sigma_j) = 2 \delta_{ij}$, we trace out both sides, to find
\begin{equation}
    \delta_{jl} = \sum_{k,m} \mathcal{R}_{kj}\mathcal{R}_{ml} \delta_{km} = \sum_k \mathcal{R}^T_{jk} \mathcal{R}_{kl},
\end{equation}
that is, $\mathcal{R}^T \mathcal{R} = \mathds{I}$. 

All in all, given a unitary $\mathcal{U} \in SU(2)$, that describes an invertible quantum channel that take a state $\rho$ to $\rho'$, we obtain a rotation $\mathcal{R} \in SO(3)$, in a way that that it describes the transition from $\vec{r}$ to $\vec{r}\,'$. Finally, this shows that the direction determined by the blue arrows in the diagram from Fig. \ref{fig:1} commutes.

Although one can argue that the correspondence between unitaries and rotations is only a two-to-one homomorphism, the situation is indeed a bijection when we restrict our discussion to invertible quantum channels acting on a qubit. Because these maps are of the form
\begin{equation}
\Phi(\cdot) = \mathcal{U}\ (\cdot )\, \mathcal{U}^*, \nonumber
\end{equation}
with $\mathcal{U} \in SU(2)$, we note that replacing $\mathcal{U}$ by $-\mathcal{U}$ yields the same quantum channel. Therefore, what
emerges is in fact an isomorphism between the group of reversible quantum maps and rotations in $\mathds{R}^3$.

We conclute this section with an alternative route that could have been taken. Our construction could equivalently begin from the direction 
$SU(2)\rightarrow SO(3)$. This alternative route is indeed very natural: the role of unitaries is already justified by the discussion of reversible CPTP maps, and such unitaries induce rotations on the Bloch sphere in a straightforward manner. Nevertheless, we adopt the opposite order (starting from transformations of physical space and then considering their induced action on the state space) because our treatment is grounded in an operational perspective. In practice, one first specifies a preparation by selecting a physical direction, and only afterwards considers how transformations act on these preparations. Moreover, both in physical space and in the space of quantum states, the set of mathematically admissible isometries is larger than the set of physically implementable transformations. To establish a meaningful group-theoretic connection between the two domains, it is therefore necessary to restrict attention to the class of transformations that correspond operationally on both sides—namely, proper rotations in $\mathds{R}^3$ and their unitary counterparts in $SU(2)$.

\section{Conclusion and Perspectives}\label{Sec.5}

In this work, we have explored a concrete physical system to operationally reconstruct the well-known two-to-one homomorphism between the groups $SU(2)$  and $SO(3)$. Rather than treating this correspondence as a purely mathematical artifact, we built it from first principles using physical procedures such as state preparation, reversible quantum evolutions, and the structure of CPTP maps. This allowed us to realize the homomorphism through a sequence of physically meaningful and experimentally motivated steps, firmly rooted in the geometry of the Bloch sphere and the algebra of quantum operations.

Our construction not only reinforces the deep interplay between group theory and quantum mechanics but also offers a clear operational standpoint through which these connections emerge. By anchoring the abstract homomorphism in transformations that can be implemented in practice—rotations in physical space and unitaries in Hilbert space—we provide a concrete interpretation that enriches the usual formal presentation.

Beyond its theoretical contribution, the approach developed here is didactically valuable. It presents a foundational concept in quantum theory in a way that is both conceptually transparent and pedagogically accessible, potentially serving as an entry door to teaching symmetries, quantum channels, and the structure of quantum state space.

We conclude our manuscript with some future directions. Our current investigation on two-level quantum systems indicates what a invertible physical map looks like in the physical space: the Bloch vector is rotated. What happens in the physical space when a general CPTP map acts on a state? It not only changes the direction of the corresponding Bloch vector, but also may rescale it, followed by a translation; see Sec. 8.3.2 of \cite{nielsen2010quantum}. We would like to investigate what is the action of an arbitrary quantum channel on states spaces of higher dimensions. For example, in the case of $\mathds{C}^4 \cong \mathds{C}^2 \otimes \mathds{C}^2$, general states are expanded as
\[
\rho = \frac{1}{4} \left( S_{00} + \sum_i r_i S_{i0} +\sum_{j}s_jS_{0j} +\sum_{i,j}t_{ij}S_{ij} \right ),
\]
where $S_{\mu \nu} := \sigma_\mu \otimes \sigma_\nu$, $\mu,\nu\in \{0,1,2,3\}$ and $S_{00} = \I \otimes \I$. 
Clearly one recognizes two vectors, $\vec{r}$, $\vec{s}$ and a second order tensor $t$, with elements $t_{ij}$ ---the correlation matrix \cite{terra-tese}. It would be interesting to see how a quantum channel act on these geometric structures representing a bipartite system. 

The connection we uncover between physical space and the state spaces of two-level quantum systems points toward several promising directions. It opens the possibility of using geometric insights to address the divisibility of quantum channels \cite{duarte_relating_2022}, and to explore the limits of channel inversion through the Petz recovery map, regarded as the quantum counterpart of Bayes’ theorem \cite{petz_sufficient_1986, leifer_towards_2013}. Beyond these, our approach may also contribute to the characterization of effective/emergent dynamics in coarse-grained systems, where the Petz map provides a natural tool. We believe that extending our results along these lines could yield a deeper and more unified understanding of how information is processed and transformed in quantum dynamics, even looking to its geometrical counterpart in the physical space.

\section{Acknowledgments}

The authors would like to thank the Brazilian agencies CAPES, CNPq and FAPEMIG for the support. 

Tha authors would also like to thank Prof. Raphael Campos Drumond, who pointed out some references on double-cover group homomorphisms. 

We are also in debt with Dr. Cristhiano Duarte, who carefully read and suggested many adjustments to the first version of this manuscript, improving both readability and content. 

\section{Appendix}
\label{sec.appendix}

Here we state and prove the corollary of Theorem \ref{theorem.1}. This is exactly the result we have used to connect invertible CPTP maps (with inverse also being CPTP) with a group structure.  

\begin{theorem}\label{theorem.1}
    Let $\Phi: L(\mathds{C}^n) \rightarrow L(\mathds{C}^n)$ be a CPTP map. Suppose that $\Phi$ admits an inverse $\Phi^{-1}$ which is also CPTP. Then there exists a unitary $\mathcal{U}$ on $\mathds{C}^n$ such that 
    $\Phi(\cdot) = \mathcal{U}(\cdot) \, \mathcal{U}^*$ and $\Phi^{-1}(\cdot) = \mathcal{U}^* (\cdot)\,\mathcal{U}$. Conversely, if $\Phi$ is a CPTP map whose Kraus representation has only one operator, say $\Phi(\cdot) = \mathcal{U}(\cdot) \, \mathcal{U}^*$, then $\Phi$ is invertible, $\Phi^{-1}(\cdot) = \mathcal{U}^* (\cdot)\,\mathcal{U}$ and $\Phi^{-1}$ is also CPTP. 
\end{theorem}
\textit{\underline{Proof:}}

Since $\Phi$ and $\Phi^{-1}$ are CPTP maps, they both can be represented via the Kraus representation—as given in equation \eqref{8.1}—using the sets of Kraus operators $\{A_a\, \vert \,a\in\mathcal{I}\}$, $\{B_b\, \vert \,b\in\mathcal{I}\} \subset L(\C^n)$:
\begin{align}   
    \Phi(\cdot)=\sum_aA_a(\cdot) A_a^*\ ; \nonumber \\\Phi^{-1}(\cdot)=\sum_bB_b(\cdot) B_b^*\ .
\end{align}
The amount of elements in the two sets of Kraus operators is a direct consequence of the fact that the cardinality of $\mathcal{I}$ is tied to the rank of the Choi-Jamiołkowski representation of the same $\Phi$ \cite{Watrous_2018}, $|\mathcal{I}| =rank(\mathrm{J}(\Phi))$, combined with the fact that $\Phi$ is invertible, resulting in $a,b\in\mathcal{I}$ having the same number of elements.

Thus we have 
\begin{equation}\label{inverse}
    \Phi^{-1}(\Phi(\cdot)) = \sum_{a,b}B_bA_a(\cdot) A_a^*B_b^*,
\end{equation}
which defines another CPTP map, namely, the identity one. Eq. \eqref{inverse} is another representation of the identity. Then, by defining $B_bA_a=\alpha_{ba}\I_n$, the complex coefficients $a,b \in \mathcal{I}$ satisfy  
$\sum_{a,b}|\alpha_{ba}|^2 = 1$, due to the unitary equivalence between Kraus representations \cite{nielsen2010quantum}.\footnote{With more details, if two sets of Kraus operators, say $\{A_a\}$ and $\{B_i\}$, for some indexes $a$ and $i$, are connected by coefficients of a unitary matrix $U$, $A_a = \sum_i U_{i a}B_i$, then they describe the same operation on the state $\rho$: 
\begin{align}
\sum_{a} A_a\rho A^*_a &= \sum_{a,i,j}U_{ia}B_i \rho \bar{U}_{ja}B_j=\sum_{i,j}(U U^*)_{ij}B_j\rho B^*_j \cr \nonumber &=\sum_{i}B_i\rho B^*_i    
\end{align}
as $(U U^*)_{ij} = \delta_{ij}$.} In fact, for any $\rho \in L(\C^n)$ we have
\begin{align}
    \sum_{a,b}B_bA_a\ \rho \ A_a^*B_b^* &= \sum_{a,b}\alpha_{ba}\I_n\ \rho \  \overline\alpha_{ba}\I_n \nonumber \\
    &=\sum_{a,b}|\alpha_{ba}|^2\rho  = \rho.
\end{align}

For convenience, we can write
\begin{equation}
\sum_bA_{a^{'}}^*B_b^*B_bA_a=\beta_{a'a}\I_n ; \ \forall a',a\in \mathcal{I},
\end{equation}
where $\beta_{a'a}=\sum_b\overline{\alpha}_{ba'}\alpha_{ba}$. 

By the Kraus representation restriction $\sum_b B_b^*B_b=\I_n$, it follows that
\begin{equation}
\label{AaAa}
    A_{a'}^*A_a=\beta_{a'a}\I_n \  ; \ \forall a',a \in \mathcal{I}.
\end{equation}
We can consider $\beta_{a'a}$ as the elements of a $a'\times a$ matrix $\mathcal{M}$ that satisfy $Tr(\mathcal{M}) = \sum_a\beta_{aa} = \sum_{ab}|\alpha_{ba}|^2=1$, and is positive semi-definite due to the fact that for every choice of $u$, $u^*\mathcal{M}u = \sum_b\left(\sum_{a'}\overline{u}_{a'}\overline{\alpha}_{ba'}\right)\left(\sum_au_a\alpha_{ba}\right)\geq 0$.

Looking at equation \eqref{AaAa}, we observe that the Kraus operators from $\Phi$ embed elements from $\C^n \ \text{to} \ \C^n$ by scaling them, in other words, they do not necessarily embed $\C^n$ into orthogonal subspaces in $\C^n$, since $\mathcal{M}$ is not necessarily diagonal. It is known that the matrix $\mathcal{M}$ is positive semi-definite, as such, it is also self-adjoint, so it allows a decomposition via the spectral theorem, meaning that $\mathcal{M}$ can be diagonalized. Therefore, to construct an equivalent representation of $\Phi$ in which the images lie in orthogonal subspaces, we must first diagonalize the matrix $\mathcal{M}$.

Given a unitary matrix $\mathcal{V}$ that diagonalizes $\mathcal{M}$, where $\Gamma=\mathcal{V}^*\mathcal{M}\mathcal{V}$ is the resulting diagonal matrix with $\gamma_a=\Gamma_{aa}\geq0 \  ; \ \forall a \in \mathcal{I}$ and $\sum_a\gamma_a=\sum_a\beta_{aa}=1$.

We now have a new Kraus operator, which we will call $C_c$ (that also represents the map $\Phi$), given by the unitary equivalence of Kraus operators as $C_c=\sum_av_{ac}A_a$, where $v_{ac}$ are the matrix elements of $\mathcal{V}$. 

For all $c$ and $c'$ in $\mathcal{I}$ we have
\begin{align}
C_{c'}^*C_c &= \left(\sum_{a'}\overline{v}_{c'a'}A_{a'}^*\right)\left(\sum_{a}v_{ca}A_{a}\right) \nonumber \\ &= \sum_{a',a}\overline{v}_{c'a'}v_{ca}(A_{a'}^*A_a).    
\end{align}

By equation \eqref{AaAa}, it follows that
\begin{align}
    C_{c'}^*C_c &= \sum_{a',a}\overline{v}_{c'a'}v_{ca}(\beta_{a'a}\I_n) \nonumber = \left(\sum_{a',a}\overline{v}_{c'a'}\beta_{a'a}v_{ca}\right)\I_n \nonumber \\ &=\Gamma_{c'c}\I_n = \delta_{c'c}\gamma_c\I_n. 
\end{align}

Defining $\mathcal{J}=\{c\in\mathcal{I} \, \vert \, \gamma_c\neq 0\}$, now we have a diagonal matrix, therefore, the Kraus operators $\{C_c \, \vert \, \forall c\in\mathcal{J}\}$ embed into orthogonal subspaces in $\C^n$, and, by singular value decomposition, $C_c$ takes the form:
\begin{equation}
    C_c=\sum_{i\in[n]} \sqrt{\gamma_c}\ket{y_{ci}}\bra{u_i^{(c)}}.
\end{equation}
where $\{y_{ci}\}_{c\in\mathcal{J}, i\in[n]}$ is an orthonormal set of vectors in $\C^n$ and $\{u_i^{(c)}\}_{i\in[n]}$ is an orthonormal basis of $\C^n$ for each $c\in\mathcal{J}$.

Therefore we have a unitary transformation $\mathcal{U}$ that, for  each $c\in\mathcal{J}$, takes the orthonormal basis $\{u_i^{(c)}\}_{i\in[n]}$ of $\C^n$ to an orthonormal set of vectors $\{y_{ci}\}_{c\in\mathcal{J}, i\in[n]}$ in $\C^n$ i.e. $\mathcal{U}: \ket{u_i^{(c)}}\mapsto\ \ket{y_{ci}}$, because $C_c=\sqrt{\gamma_c} \ \mathcal{U}$ via the  the unitary equivalence of Kraus operators.

Thus, 
 we have
\begin{align}
    \Phi(\rho) &= \sum_cC_c\, \rho \ C_c^* \nonumber \\
    &= \sum_c\gamma_c\ \mathcal{U}\, \rho \ \mathcal{U} 
    =\mathcal{U} \,\rho\ \mathcal{U}.
\end{align}
The unitary matrix $\mathcal{U}$ represents the action of all the Kraus operators $C_c$ on $\rho$ for every $c\in\mathcal{J}$, reducing the Kraus representation of $\Phi$ to one single element. That allows us to write $\Phi$ as $\Phi(\cdot)=\mathcal{U}\,(\cdot)\,\mathcal{U}^*$, and since $\Phi^{-1}(\Phi(\cdot))$ represents the identity map, it follows that $\Phi^{-1}(\cdot)=\mathcal{U}^*(\cdot)\,\mathcal{U}$. Thus completing the first part of the proof.

The converse part of this theorem follows from a straightforward sequence of steps. A map is said to be invertible if there exists another map $\Phi^{-1}$ such that applying $\Phi^{-1}$ followed by $\Phi$ (or vice versa) yields the identity map. Consider the map defined by $\Phi^{-1}(\cdot) = \mathcal{U}^* (\cdot) \,  \mathcal{U}$. This map effectively undoes the action of $\Phi$. 

Given an arbitrary state $\sigma$
\begin{equation}
    \Phi^{-1}(\Phi(\sigma)) = \mathcal{U}^* \mathcal{U}\, \sigma\, \mathcal{U}^* \mathcal{U} = \sigma \Rightarrow \Phi^{-1}\circ \Phi = \I_n 
\end{equation}
since $\mathcal{U}$ is unitary. 

As shown in equations \eqref{8.1} and \eqref{8.11}, $\Phi^{-1}$ has the structure of a CPTP map. Therefore, $\Phi$ admits an inverse that is also CPTP, completing the proof of the converse. $\qedsymbol{}$


\begin{thebibliography}{23}%
	\makeatletter
	\providecommand \@ifxundefined [1]{%
		\@ifx{#1\undefined}
	}%
	\providecommand \@ifnum [1]{%
		\ifnum #1\expandafter \@firstoftwo
		\else \expandafter \@secondoftwo
		\fi
	}%
	\providecommand \@ifx [1]{%
		\ifx #1\expandafter \@firstoftwo
		\else \expandafter \@secondoftwo
		\fi
	}%
	\providecommand \natexlab [1]{#1}%
	\providecommand \enquote  [1]{``#1''}%
	\providecommand \bibnamefont  [1]{#1}%
	\providecommand \bibfnamefont [1]{#1}%
	\providecommand \citenamefont [1]{#1}%
	\providecommand \href@noop [0]{\@secondoftwo}%
	\providecommand \href [0]{\begingroup \@sanitize@url \@href}%
	\providecommand \@href[1]{\@@startlink{#1}\@@href}%
	\providecommand \@@href[1]{\endgroup#1\@@endlink}%
	\providecommand \@sanitize@url [0]{\catcode `\\12\catcode `\$12\catcode
		`\&12\catcode `\#12\catcode `\^12\catcode `\_12\catcode `\%12\relax}%
	\providecommand \@@startlink[1]{}%
	\providecommand \@@endlink[0]{}%
	\providecommand \url  [0]{\begingroup\@sanitize@url \@url }%
	\providecommand \@url [1]{\endgroup\@href {#1}{\urlprefix }}%
	\providecommand \urlprefix  [0]{URL }%
	\providecommand \Eprint [0]{\href }%
	\providecommand \doibase [0]{https://doi.org/}%
	\providecommand \selectlanguage [0]{\@gobble}%
	\providecommand \bibinfo  [0]{\@secondoftwo}%
	\providecommand \bibfield  [0]{\@secondoftwo}%
	\providecommand \translation [1]{[#1]}%
	\providecommand \BibitemOpen [0]{}%
	\providecommand \bibitemStop [0]{}%
	\providecommand \bibitemNoStop [0]{.\EOS\space}%
	\providecommand \EOS [0]{\spacefactor3000\relax}%
	\providecommand \BibitemShut  [1]{\csname bibitem#1\endcsname}%
	\let\auto@bib@innerbib\@empty
	\bibitem [{\citenamefont {Leifer}\ and\ \citenamefont
		{Spekkens}(2013)}]{leifer_towards_2013}%
	\BibitemOpen
	\bibfield  {author} {\bibinfo {author} {\bibfnamefont {M.~S.}\ \bibnamefont
			{Leifer}}\ and\ \bibinfo {author} {\bibfnamefont {R.~W.}\ \bibnamefont
			{Spekkens}},\ }\bibfield  {title} {\bibinfo {title} {Towards a formulation of
			quantum theory as a causally neutral theory of {Bayesian} inference},\
	}\href@noop {} {\bibfield  {journal} {\bibinfo  {journal} {Physical Review
				A}\ }\textbf {\bibinfo {volume} {88}},\ \bibinfo {pages} {052130} (\bibinfo
		{year} {2013})}\BibitemShut {NoStop}%
	\bibitem [{\citenamefont {Wolf}(2012)}]{wolf}%
	\BibitemOpen
	\bibfield  {author} {\bibinfo {author} {\bibfnamefont {M.~M.}\ \bibnamefont
			{Wolf}},\ }\bibfield  {title} {\bibinfo {title} {Quantum channels and
			operations: guided tour},\ }\href@noop {} {\bibfield  {journal} {\bibinfo
			{journal} {Lecture notes available at url =
				{https://mediatum.ub.tum.de/doc/1701036/1701036.pdf}}\ } (\bibinfo {year}
		{2012})}\BibitemShut {NoStop}%
	\bibitem [{\citenamefont {Duarte}\ \emph {et~al.}(2017)\citenamefont {Duarte},
		\citenamefont {Carvalho}, \citenamefont {Bernardes},\ and\ \citenamefont
		{de~Melo}}]{duarte_emerging_2017}%
	\BibitemOpen
	\bibfield  {author} {\bibinfo {author} {\bibfnamefont {C.}~\bibnamefont
			{Duarte}}, \bibinfo {author} {\bibfnamefont {G.~D.}\ \bibnamefont
			{Carvalho}}, \bibinfo {author} {\bibfnamefont {N.~K.}\ \bibnamefont
			{Bernardes}},\ and\ \bibinfo {author} {\bibfnamefont {F.}~\bibnamefont
			{de~Melo}},\ }\bibfield  {title} {\bibinfo {title} {Emerging dynamics arising
			from coarse-grained quantum systems},\ }\href@noop {} {\bibfield  {journal}
		{\bibinfo  {journal} {Physical Review A}\ }\textbf {\bibinfo {volume} {96}},\
		\bibinfo {pages} {032113} (\bibinfo {year} {2017})}\BibitemShut {NoStop}%
	\bibitem [{\citenamefont {Horodecki}\ \emph {et~al.}(2009)\citenamefont
		{Horodecki}, \citenamefont {Horodecki}, \citenamefont {Horodecki},\ and\
		\citenamefont {Horodecki}}]{horodecki_quantum_2009}%
	\BibitemOpen
	\bibfield  {author} {\bibinfo {author} {\bibfnamefont {R.}~\bibnamefont
			{Horodecki}}, \bibinfo {author} {\bibfnamefont {P.}~\bibnamefont
			{Horodecki}}, \bibinfo {author} {\bibfnamefont {M.}~\bibnamefont
			{Horodecki}},\ and\ \bibinfo {author} {\bibfnamefont {K.}~\bibnamefont
			{Horodecki}},\ }\bibfield  {title} {\bibinfo {title} {Quantum entanglement},\
	}\href@noop {} {\bibfield  {journal} {\bibinfo  {journal} {Reviews of Modern
				Physics}\ }\textbf {\bibinfo {volume} {81}},\ \bibinfo {pages} {865}
		(\bibinfo {year} {2009})}\BibitemShut {NoStop}%
	\bibitem [{\citenamefont {Marquardt}\ and\ \citenamefont
		{Püttmann}(2008)}]{marquardt_introduction_2008}%
	\BibitemOpen
	\bibfield  {author} {\bibinfo {author} {\bibfnamefont {F.}~\bibnamefont
			{Marquardt}}\ and\ \bibinfo {author} {\bibfnamefont {A.}~\bibnamefont
			{Püttmann}},\ }\bibfield  {title} {\bibinfo {title} {Introduction to
			dissipation and decoherence in quantum systems},\ }\href@noop {} {\bibfield
		{journal} {\bibinfo  {journal} {arXiv:0809.4403 [quant-ph]}  \bibinfo
			{pages} {}} (\bibinfo {year} {2008})}\BibitemShut {NoStop}%
	\bibitem [{\citenamefont {Plávala}(2023)}]{plavala_general_2023}%
	\BibitemOpen
	\bibfield  {author} {\bibinfo {author} {\bibfnamefont {M.}~\bibnamefont
			{Plávala}},\ }\bibfield  {title} {\bibinfo {title} {General probabilistic
			theories: {An} introduction},\ }\href@noop {} {\bibfield  {journal} {\bibinfo
			{journal} {Physics Reports}\ }\textbf {\bibinfo {volume} {1033}},\ \bibinfo
		{pages} {1} (\bibinfo {year} {2023})}\BibitemShut {NoStop}%
	\bibitem [{\citenamefont {Nielsen}\ and\ \citenamefont
		{Chuang}(2010)}]{nielsen2010quantum}%
	\BibitemOpen
	\bibfield  {author} {\bibinfo {author} {\bibfnamefont {M.~A.}\ \bibnamefont
			{Nielsen}}\ and\ \bibinfo {author} {\bibfnamefont {I.~L.}\ \bibnamefont
			{Chuang}},\ }\href@noop {} {\emph {\bibinfo {title} {Quantum Computation and
				Quantum Information}}}\ (\bibinfo  {publisher} {Cambridge University Press},\
	\bibinfo {year} {2010})\BibitemShut {NoStop}%
	\bibitem [{\citenamefont {Nayak}\ and\ \citenamefont
		{Sen}(2007)}]{nayak2006invertible}%
	\BibitemOpen
	\bibfield  {author} {\bibinfo {author} {\bibfnamefont {A.}~\bibnamefont
			{Nayak}}\ and\ \bibinfo {author} {\bibfnamefont {P.}~\bibnamefont {Sen}},\
	}\bibfield  {title} {\bibinfo {title} {Invertible quantum operations and
			perfect encryption of quantum states},\ }\href@noop {} {\bibfield  {journal}
		{\bibinfo  {journal} {Quantum Info. Comput.}\ }\textbf {\bibinfo {volume}
			{7}},\ \bibinfo {pages} {103–110} (\bibinfo {year} {2007})}\BibitemShut
	{NoStop}%
	\bibitem [{\citenamefont {Valle}\ \emph {et~al.}(2024)\citenamefont {Valle},
		\citenamefont {Brugger}, \citenamefont {Rizzuti},\ and\ \citenamefont
		{Duarte}}]{valle2024towards}%
	\BibitemOpen
	\bibfield  {author} {\bibinfo {author} {\bibfnamefont {V.}~\bibnamefont
			{Valle}}, \bibinfo {author} {\bibfnamefont {L.}~\bibnamefont {Brugger}},
		\bibinfo {author} {\bibfnamefont {B.}~\bibnamefont {Rizzuti}},\ and\ \bibinfo
		{author} {\bibfnamefont {C.}~\bibnamefont {Duarte}},\ }\bibfield  {title}
	{\bibinfo {title} {Towards establishing a connection between two-level
			quantum systems and physical spaces},\ }\href@noop {} {\bibfield  {journal}
		{\bibinfo  {journal} {Brazilian Journal of Physics}\ }\textbf {\bibinfo
			{volume} {54}},\ \bibinfo {pages} {93} (\bibinfo {year} {2024})}\BibitemShut
	{NoStop}%
	\bibitem [{\citenamefont {Peres}(2010)}]{peres_quantum_2010}%
	\BibitemOpen
	\bibfield  {author} {\bibinfo {author} {\bibfnamefont {A.}~\bibnamefont
			{Peres}},\ }\href@noop {} {\emph {\bibinfo {title} {Quantum theory: concepts
				and methods}}} 
	(\bibinfo  {publisher} {Kluwer Acad. Publ},\ \bibinfo {address} {Dordrecht},\
	\bibinfo {year} {2010})\BibitemShut {NoStop}%
	\bibitem [{\citenamefont {Gaio}\ \emph {et~al.}(2019)\citenamefont {Gaio},
		\citenamefont {de~Barros},\ and\ \citenamefont
		{Rizzuti}}]{gaio_grandezas_2019}%
	\BibitemOpen
	\bibfield  {author} {\bibinfo {author} {\bibfnamefont {L.~M.}\ \bibnamefont
			{Gaio}}, \bibinfo {author} {\bibfnamefont {D.~R.~T.}\ \bibnamefont
			{de~Barros}},\ and\ \bibinfo {author} {\bibfnamefont {B.~F.}\ \bibnamefont
			{Rizzuti}},\ }\bibfield  {title} {\bibinfo {title} {Grandezas físicas
			multidimensionais},\ }\href@noop {} {\bibfield  {journal} {\bibinfo
			{journal} {Revista Brasileira de Ensino de Física}\ }\textbf {\bibinfo
			{volume} {41}},\ \bibinfo {pages} {e20180295} (\bibinfo {year}
		{2019})}\BibitemShut {NoStop}%
	\bibitem [{\citenamefont {Rizzuti}\ \emph {et~al.}(2020)\citenamefont
		{Rizzuti}, \citenamefont {Gaio},\ and\ \citenamefont
		{Duarte}}]{rizzuti_operational_2020}%
	\BibitemOpen
	\bibfield  {author} {\bibinfo {author} {\bibfnamefont {B.~F.}\ \bibnamefont
			{Rizzuti}}, \bibinfo {author} {\bibfnamefont {L.~M.}\ \bibnamefont {Gaio}},\
		and\ \bibinfo {author} {\bibfnamefont {C.}~\bibnamefont {Duarte}},\
	}\bibfield  {title} {\bibinfo {title} {Operational {Approach} to the
			{Topological} {Structure} of the {Physical} {Space}},\ }\href@noop {}
	{\bibfield  {journal} {\bibinfo  {journal} {Foundations of Science}\ }\textbf
		{\bibinfo {volume} {25}},\ \bibinfo {pages} {711} (\bibinfo {year}
		{2020})}\BibitemShut {NoStop}%
	\bibitem [{\citenamefont {Grossi}\ \emph {et~al.}(2023)\citenamefont {Grossi},
		\citenamefont {Brugger}, \citenamefont {Rizzuti},\ and\ \citenamefont
		{Duarte}}]{grossi2023one}%
	\BibitemOpen
	\bibfield  {author} {\bibinfo {author} {\bibfnamefont {R.}~\bibnamefont
			{Grossi}}, \bibinfo {author} {\bibfnamefont {L.~L.}\ \bibnamefont {Brugger}},
		\bibinfo {author} {\bibfnamefont {B.}~\bibnamefont {Rizzuti}},\ and\ \bibinfo
		{author} {\bibfnamefont {C.}~\bibnamefont {Duarte}},\ }\bibfield  {title}
	{\bibinfo {title} {One hundred years later: Stern-gerlach experiment and
			dimension witnesses},\ }\href@noop {} {\bibfield  {journal} {\bibinfo
			{journal} {Revista Brasileira de Ensino de F{\'\i}sica}\ }\textbf {\bibinfo
			{volume} {45}},\ \bibinfo {pages} {e20220227} (\bibinfo {year}
		{2023})}\BibitemShut {NoStop}%
	\bibitem [{\citenamefont {Bengtsson}\ and\ \citenamefont
		{{\.Z}yczkowski}(2017)}]{bengtsson2017geometry}%
	\BibitemOpen
	\bibfield  {author} {\bibinfo {author} {\bibfnamefont {I.}~\bibnamefont
			{Bengtsson}}\ and\ \bibinfo {author} {\bibfnamefont {K.}~\bibnamefont
			{{\.Z}yczkowski}},\ }\href@noop {} {\emph {\bibinfo {title} {Geometry of
				Quantum States: An Introduction to Quantum Entanglement}}}\ (\bibinfo
	{publisher} {Cambridge University Press},\ \bibinfo {year}
	{2017})\BibitemShut {NoStop}%
	\bibitem [{\citenamefont {Gamel}(2016)}]{gamel_entangled_2016}%
	\BibitemOpen
	\bibfield  {author} {\bibinfo {author} {\bibfnamefont {O.}~\bibnamefont
			{Gamel}},\ }\bibfield  {title} {\bibinfo {title} {Entangled {Bloch} spheres:
			{Bloch} matrix and two-qubit state space},\ }\href@noop {} {\bibfield
		{journal} {\bibinfo  {journal} {Physical Review A}\ }\textbf {\bibinfo
			{volume} {93}},\ \bibinfo {pages} {062320} (\bibinfo {year}
		{2016})}\BibitemShut {NoStop}%
	\bibitem [{\citenamefont {Mukunda}\ \emph {et~al.}(2010)\citenamefont
		{Mukunda}, \citenamefont {Chaturvedi},\ and\ \citenamefont
		{Simon}}]{mukunda2010hamilton}%
	\BibitemOpen
	\bibfield  {author} {\bibinfo {author} {\bibfnamefont {N.}~\bibnamefont
			{Mukunda}}, \bibinfo {author} {\bibfnamefont {S.}~\bibnamefont
			{Chaturvedi}},\ and\ \bibinfo {author} {\bibfnamefont {R.}~\bibnamefont
			{Simon}},\ }\bibfield  {title} {\bibinfo {title} {Hamilton’s theory of
			turns revisited},\ }\href@noop {} {\bibfield  {journal} {\bibinfo  {journal}
			{Pramana}\ }\textbf {\bibinfo {volume} {74}},\ \bibinfo {pages} {1} (\bibinfo
		{year} {2010})}\BibitemShut {NoStop}%
	\bibitem [{\citenamefont {Tung}(1985)}]{tung_group_1985}%
	\BibitemOpen
	\bibfield  {author} {\bibinfo {author} {\bibfnamefont {W.-K.}\ \bibnamefont
			{Tung}},\ }\href@noop {} {\emph {\bibinfo {title} {Group Theory in
				Physics}}}\ (\bibinfo  {publisher} {World Scientific},\ \bibinfo {address}
	{Philadelphia},\ \bibinfo {year} {1985})\BibitemShut {NoStop}%
	\bibitem [{\citenamefont {Watrous}(2018)}]{Watrous_2018}%
	\BibitemOpen
	\bibfield  {author} {\bibinfo {author} {\bibfnamefont {J.}~\bibnamefont
			{Watrous}},\ }\href@noop {} {\emph {\bibinfo {title} {The Theory of Quantum
				Information}}}\ (\bibinfo  {publisher} {Cambridge University Press},\
	\bibinfo {year} {2018})\BibitemShut {NoStop}%
	\bibitem [{\citenamefont {Hall}(2015)}]{hall_lie_2015}%
	\BibitemOpen
	\bibfield  {author} {\bibinfo {author} {\bibfnamefont {B.~C.}\ \bibnamefont
			{Hall}},\ }\href@noop {} {\emph {\bibinfo {title} {Lie {Groups}, {Lie}
				{Algebras}, and {Representations}: {An} {Elementary} {Introduction}}}},\
	Graduate {Texts} in {Mathematics}\ (\bibinfo  {publisher} {Springer
		International Publishing},\ \bibinfo {address} {Cham},\ \bibinfo {year}
	{2015})\BibitemShut {NoStop}%
	\bibitem [{\citenamefont {Rocha}\ \emph {et~al.}(2013)\citenamefont {Rocha},
		\citenamefont {Rizzuti},\ and\ \citenamefont
		{Mota}}]{rocha_transformacoes_2013}%
	\BibitemOpen
	\bibfield  {author} {\bibinfo {author} {\bibfnamefont {A.}~\bibnamefont
			{Rocha}}, \bibinfo {author} {\bibfnamefont {B.}~\bibnamefont {Rizzuti}},\
		and\ \bibinfo {author} {\bibfnamefont {D.}~\bibnamefont {Mota}},\ }\bibfield
	{title} {\bibinfo {title} {Transformações de {Galileu} e de {Lorentz}: um
			estudo via teoria de grupos},\ }\href@noop {} {\bibfield  {journal} {\bibinfo
			{journal} {Revista Brasileira de Ensino de Física}\ }\textbf {\bibinfo
			{volume} {35}} (\bibinfo {year} {2013})}\BibitemShut {NoStop}%
	\bibitem [{\citenamefont {de~O.~Terra~Cunha}(2005)}]{terra-tese}%
	\BibitemOpen
	\bibfield  {author} {\bibinfo {author} {\bibfnamefont {M.}~\bibnamefont
			{de~O.~Terra~Cunha}},\ }\href@noop {} {\emph {\bibinfo {title}
			{Emaranhamento: caracterização, manipulação e consequências}}}\
	(\bibinfo  {publisher} {Tese de doutorado - UFMG},\ \bibinfo {address} {Belo
		Horizonte},\ \bibinfo {year} {2005})\BibitemShut {NoStop}%
	\bibitem [{\citenamefont {Duarte}\ \emph {et~al.}(2022)\citenamefont {Duarte},
		\citenamefont {Catani},\ and\ \citenamefont
		{Drumond}}]{duarte_relating_2022}%
	\BibitemOpen
	\bibfield  {author} {\bibinfo {author} {\bibfnamefont {C.}~\bibnamefont
			{Duarte}}, \bibinfo {author} {\bibfnamefont {L.}~\bibnamefont {Catani}},\
		and\ \bibinfo {author} {\bibfnamefont {R.~C.}\ \bibnamefont {Drumond}},\
	}\bibfield  {title} {\bibinfo {title} {Relating {Compatibility} and
			{Divisibility} of {Quantum} {Channels}},\ }\href@noop {} {\bibfield
		{journal} {\bibinfo  {journal} {International Journal of Theoretical
				Physics}\ }\textbf {\bibinfo {volume} {61}},\ \bibinfo {pages} {189}
		(\bibinfo {year} {2022})}\BibitemShut {NoStop}%
	\bibitem [{\citenamefont {Petz}(1986)}]{petz_sufficient_1986}%
	\BibitemOpen
	\bibfield  {author} {\bibinfo {author} {\bibfnamefont {D.}~\bibnamefont
			{Petz}},\ }\bibfield  {title} {\bibinfo {title} {Sufficient subalgebras and
			the relative entropy of states of a von {Neumann} algebra},\ }\href@noop {}
	{\bibfield  {journal} {\bibinfo  {journal} {Communications in Mathematical
				Physics}\ }\textbf {\bibinfo {volume} {105}},\ \bibinfo {pages} {123}
		(\bibinfo {year} {1986})}\BibitemShut {NoStop}%
\end{thebibliography}

%

\end{document}